\pdfoutput=1
\documentclass[12pt,a4paper]{article}
\usepackage{lineno}
\usepackage{caption}
\usepackage[bookmarks=true,colorlinks=true,citecolor=blue,linkcolor=magenta,urlcolor=cyan]{hyperref}
\usepackage{ulem}
\usepackage[super]{natbib}

\usepackage{graphicx}
\usepackage{natbib}
\usepackage{amsmath}
\usepackage{amssymb}
\usepackage{subfigure}
\usepackage{url}
\usepackage{CJK}

\newcommand{\aap}{    {\it Astron. Astrophys.}}

\newcommand{\aapr}{   {\it Astron. Astrophys. Rev.}}

\newcommand{\apj}{    {\it Astrophys. J.}}

\newcommand{\apjl}{   {\it Astrophys. J. Lett.}}

\newcommand{\jgr}{    {\it J. Geophys. Res.}}

\newcommand{\nat}{    {\it Nature}}

\newcommand{\solphys}{{\it Solar Phys.}}

\renewcommand{\vec}[1]{ {\mathbf #1} }

\newcommand{\curl}{ {\bf \nabla} \times}

\newcommand{\Fig}{{Fig.}}

\begin{document}
\title{\bf Decipher the Three-Dimensional Magnetic Topology of a Great Solar Flare}
\maketitle

{\Large  \bigskip{}
 \bigskip{}
 }{\Large \par}

 \author{Chaowei Jiang$^{1,2,3,\ast}$, Peng Zou$^{1}$, Xueshang Feng$^{2,1}$, Qiang Hu$^{3}$, Aiying Duan$^{4},$
   Pingbing Zuo$^{1}$, Yi Wang$^{1}$, Fengsi Wei$^{1}$\\
   \\
   {$^{1}$Institute of Space Science and Applied Technology,
  Harbin Institute of Technology, Shenzhen 518055, China}\\
   \normalsize{$^{2}$SIGMA Weather Group, State Key Laboratory for
     Space Weather, National Space Science Center, Chinese
     Academy of Sciences, Beijing 100190, China}\\
   \normalsize{$^{3}$Center for Space Plasma \& Aeronomic Research,
   The University of Alabama in Huntsville, Huntsville, AL 35899, USA}\\
   \normalsize{$^{4}${Key Laboratory of Computational Geodynamics, University
  of Chinese Academy of Sciences, Beijing 100049, China}
   \\

   \normalsize{$^\ast$Correspondence author; E-mail: chaowei@hit.edu.cn}
  }

\newpage



\textbf{
Three-dimensional magnetic topology of solar flare plays a crucial role in understanding its explosive release of magnetic energy in the corona. However,
such three-dimensional coronal magnetic field is still elusive in direct observation. Here we realistically simulate the magnetic evolution during the eruptive process of a great flare, using a numerical magnetohydrodynamic model constrained by observed solar vector magnetogram. The numerical results reveal that the pre-flare corona contains multi-set twisted magnetic flux, which forms a coherent rope during the eruption. The rising flux rope is wrapped by a quasi-separatrix layer, which intersects itself below the rope, forming a hyperbolic flux tube and magnetic reconnection is triggered there. By tracing the footprint of the newly-reconnected field lines, we reproduce both the spatial location and its temporal evolution of flare ribbons with an expected accuracy in comparison of observed images. This scenario strongly confirms the three-dimensional version of standard flare model.
  }

\bigskip{}


Magnetic field plays a fundamental role in many astrophysical
activities, especially, solar flares, which are among the most
energetic events from the Sun. Often accompanied with coronal
mass ejections (CMEs)~\citep{Schmieder2013}, solar flares 
are commonly believed to be powered by free magnetic energy in the solar corona, a plasma
environment dominated by magnetic field~\citep{Aschwanden2004Book}. Magnetic
reconnection is thought to be the central mechanism that converts
free magnetic energy into radiation, energetic particle acceleration,
and kinetic energy of plasma~\citep{Priest2002}. Thus unraveling the
magnetic topology responsible for magnetic reconnection is
essential for understanding the nature of solar flares. However,
without measurements of the three-dimensional (3D) magnetic field in
the corona,
it is still a long-standing question: what is the magnetic
configuration and evolution of solar flares?

The macroscopic physical behavior of the solar corona can be described
in the first principle by magnetohydrodynamics (MHD). Within the MHD framework, many
theoretical models for solar flare have been proposed~\citep{Shibata2011}, either
analytically or numerically. The most commonly
invoked one is the so-called standard flare model, initially developed in
two-dimensions~\cite{Carmichael1964, Sturrock1966, Hirayama1974,
  Kopp1976} and recently extended to 3D~\cite{Aulanier2010,
  Aulanier2012, Janvier2014, Janvier2015}. It mainly concerns a
magnetically bipolar source region, the simplest form of solar active
regions (ARs), which embeds a set of twisted field lines, i.e.,
magnetic flux rope (MFR). Catastrophic loss of equilibrium of the MFR
system could be induced by certain kind of instability, such as the
kink instability (KI), which depends on the twist degree of the field
line in the rope~\citep{Hood1981, Torok2004, Fan2003}, and the torus
instability (TI), which depends on the strength of the field that
straps the rope~\citep{Bateman1978, Kliem2006PRL}. In the wake of the rising MFR
which eventually develops to CME, an electric current sheet (CS)
forms between the stretched magnetic field lines tethering the
MFR. In full 3D, the CS is formed in a topological quasi-separatrix
layer (QSL) that wraps up the MFR, and reconnection occurs mainly
below the rope, between the sheared arcades in a tether-cutting
form~\citep{Moore2006}. The observed ribbons of brightening at the
chromosphere, i.e., flare ribbons, are among the most representative
features of flares and are thought to be the footpoints of the
reconnection field lines.




It is challenging to fully fit the theoretical models into the
realistic solar flares. Although they have been widely used to partially interpret
various observed manifestations of flares in plasma emissions, the key parameter,
 i.e., the 3D distribution and temporal evolution of the coronal magnetic field underlying the
activities, is still prohibitive to obtain from observations.
Furthermore, the realistic flare eruptions are often much more
complex
in both the
pre-eruptive magnetic configuration and eruption process, which can be inferred from the complexity
of the magnetic flux distribution as measured on the solar surface, i.e., the photosphere. Thus, to decipher these flares is beyond the capability of the idealized and theoretical models.

Here we decipher the 3D magnetic configuration of a great eruptive flare
through numerical MHD simulation initialized by pre-flare coronal
field reconstruction, which realistically reproduces the dynamic
evolution of the magnetic field underlying the flare. The studied
flare, reaching GOES X9.3 class, was the largest one of the last
decade, and quickly drew intensive attentions~\citep{YangS2017, SunX2017, Warren2017, YanX2018, LiY2018, WangH2018, HuangN2018} in the communities of solar physics as well as space weather.
The flare occurred in a magnetic complex due to the
interaction of multiple magnetic polarities as observed on the
photosphere. In the following we show that immediately prior to the X9.3
flare, there exists an unstable MFR possibly due to the TI, and the
eruption is resulted by the consequent expansion of the MFR, which
follows the basic picture of the standard flare model in 3D analogy.


\section*{Results}
\subsection*{Overview of the event.}
The investigated flare took place in a super flare-productive solar AR, NOAA
12673, which ranks the first in the solar cycle 24.
It produced 4 X-class and 27 M-class flares from 2017 September 4 to 10. The X9.3 flare on September 6
started at 11:53~UT, impulsively reached its peak at 12:02~UT and then
ended at 12:10~UT. Its location on the solar disk
is shown in \Fig~\ref{fig1}, as imaged by the Atmospheric Imaging
Assembly (AIA) onboard the Solar Dynamics Observatory (SDO). When this
AR rotated to the solar limb on 2017 September~10 (\Fig~\ref{fig1}c),
it produced an X8.2 flare, which is the second largest one after the
X9.3 flare. As the two flares are generated in the same region, they
are likely to have basically the similar 3D magnetic
configuration. Thus the observation of this limb flare provides a side
view of the 3D structure underlying the flares, in addition to the
nearly top view for the X9.3 flare.

\subsection*{Magnetic field on the photosphere.}
The magnetic field on the photosphere is measured by Helioseismic and Magnetic
Imager (HMI) onboard SDO, which
provides the input to our numerical models (see Methods). \Fig~\ref{fig1}d shows a
single snapshot of the photospheric magnetic field, i.e., a pre-flare
vector magnetogram observed at 11:36~UT, just 17~min ahead of the
flare onset. There are mainly 4 magnetic polarities as seen in the
magnetic flux map. Two closely touched polarities, P0 and N0,
separated by a polarity inversion line (PIL) of C shape (referred to as the main PIL hereafter), are
located in the core region, which is nearly enclosed by another two
polarities (P1 and N1) in the south and north, respectively. Analysis
of the time-sequence magnetograms suggested that such configuration is
formed by the blocking effect of a pre-existing sunspot to several
groups of extremely fast emerging flux~\citep{YangS2017}, which results in a
strongly distorted magnetic system. Significant magnetic shear can be
seen along the main PIL, which is so strong that magnetic bald patches
(BPs)~\cite{Titov1993} form on almost the whole PIL (\Fig~\ref{fig1}e). The presence of
BPs means that magnetic field lines immediately above the PIL do not connect P0-N0
directly but are concave upward, grazing over the PIL and forming magnetic dips. Such magnetic-sheared
configuration with BPs is found in the case of theoretical models of
coronal MFR that is partially attached at the
photosphere~\citep{GibsonFan2006,
  Aulanier2010}.
Furthermore, distribution of strong electric currents with inverse
directions on two sides of the main PIL is derived from the transverse
components of the magnetic field (\Fig~\ref{fig1}f), which indicates
that volumetric current channels through in the
corona~\cite{Janvier2014, SunX2015}.
The current is significantly non-neutralized in magnetic flux of either sign, which is recognized to be a common feature of eruptive ARs~\citep{LiuY2017, Kontogiannis2017}. All these features suggest that a twisted MFR exists in the region and can likely account for the eruptive activities.

\subsection*{Pre-flare Magnetic Configuration.}
The coronal magnetic field in a quasi-static state prior to flare can
be well approximated by force-free models~\citep{Wiegelmann2008}. From the HMI vector magnetogram, our NLFFF
model~\citep{Jiang2013NLFFF} reconstructs magnetic field lines that
nicely match the observed coronal loops (see \Fig~\ref{fig2}). The
pre-flare magnetic configuration consists of a set of strongly-sheared
(current-carrying) low-lying (heights of $\sim 10$~Mm) field lines in
the core region, which is enveloped by less sheared
arcades (\Fig~\ref{fig2}a and d). The low-lying field lines extend their ends to the south and
north polarities P1 and N1, forming an overall C shape. Magnetic dips
are found along these field lines, and solar filament can be supported
in the dips by the upward magnetic tension force against the downward
solar gravity. As shown in \Fig~\ref{fig2}e, the dips distribute
almost all the way along the main PIL, which is in consistence with
the distribution of BPs. They can thus support a long filament along
the PIL. Such filament appears to exist as seen in the GONG H$\alpha$
image (\Fig~\ref{fig2}f), and the shape of the dips matches rather well with the
filament.

The existence of MFR is confirmed by the NLFFF model. Calculation of
the magnetic twist number $T_w$, i.e., number of turns two
infinitesimally close field lines wind about each
other~\citep{Inoue2011, LiuR2016}, shows that the low-lying core field
is twisted left-handedly (see \Fig~\ref{fig3}). The magnetic twist
number ($T_w < 0$) is significantly enhanced along the two sides of
the main PIL, nearly in the same locations of intense
current density (compare \Fig~\ref{fig1}f and \Fig~\ref{fig3}a). The
regions of enhanced negative twist also extend to the far-side
polarities P1 and N1. A major part of the magnetic twist number in
these regions is $|T_w|=1.5$, which is close to the threshold
of KI for an idealized MFR~\citep{Torok2004}. Here the twisted
magnetic flux constitutes a complex flux-rope configuration with
multiple field-line connections.  The different connections can be
distinguished from a map of magnetic squashing degree (the $Q$
factor)~\citep{Demoulin1996, Titov2002}, which precisely locates the
topological separatries or quasi-separatrix layers (QSLs) of the
field-line connectivity. As shown in \Fig~\ref{fig3}b, this twisted
flux bundle consists of mainly three types of connection: P1-N0, P0-N1
and P1-N1, while connection of P0-N0 does not form along the main PIL,
a natural result of the existence of the BPs. Notably, the MFR is
strongly non-uniformly twisted, as part of field lines connecting P1
and N1 is only weakly twisted ($|T_w|<0.5$). Such multiple connections
and inhomogeneous twist are not characterized by any current
theoretical (or idealized) models of MFR.

We further compute the decay index $n$ of the strapping field, which
is the key parameter deciding the TI of an MFR system.
It is found that the major part of the MFR (i.e.,
the flux with $|T_w| > 1$) reaches a region with $n>1.5$~(see
\Fig~\ref{fig3}c), which is a
theoretical threshold of TI according to previous
studies~\citep{Kliem2006PRL, Aulanier2010}.  This indicates that the
MFR is already in an unstable regime, suggesting that the eruption is
more likely a result of TI rather than KI.
\subsection*{The Eruptive Evolution.}
The onset of the flare is characterized by a drastic rise and
expansion of the MFR in the MHD simulation, which is shown in
\Fig~\ref{fig4} for a vertical cross section and \Fig~\ref{fig5}
for the 3D configuration. As clearly seen from the cross
section of current density (\Fig~\ref{fig4}a), there forms a narrow
current layer of upside-down teardrop shape, which actually represents
the closed boundary layer of the MFR. Such boundary is precisely
depicted by the QSLs that separate the MFR with its surrounding flux, as
shown in the map of magnetic squashing degree
(\Fig~\ref{fig4}c). Below the MFR, a more intense CS forms connecting
the cusp of the teardrop to the bottom. Interestingly, there is an
intersection of QSLs below the MFR, which forms an X shape of
increasing height and size. Such intersection of QSLs is a magnetic
null point configuration in the 2D plane, while in 3D it is a
hyperbolic flux tube (HFT)~\citep{Titov2002} which possesses the
highest values of $Q$. Strong current density preferentially forms
there and triggers reconnection. Indeed, the CS also evolves to an X
shape since magnetic reconnection is triggered in the HFT, leaving
another cusp connecting the bottom boundary.

The plasma flows near the
CS are plotted in \Fig~\ref{fig6}a, which shows a typical pattern of
reconnection flow in 2D, i.e., inflows at two sides of the CS and
outflows away from its two ends. This reconnection along with the
cusp-CS-rope configuration (see the 2D field lines in
\Fig~\ref{fig6}a) reproduce nicely the picture of the standard flare
model in 2D. Furthermore, the shape of enhanced current layer
and its evolution look rather similar to the AIA images of the limb
X8.2 flare of the same AR~(\Fig~\ref{fig4}b), which show a bright ring enclosing
a relatively dark cavity of increasing size. The coronal cavity often
indicates an MFR~\citep{GibsonFan2006}, while its outer edge is bright because heating is
enhanced there by the dissipation of the strong current in the boundary layer of
rope. Below the cavity is an even brighter cusp connecting the solar surface, which is also matched by
the simulation.
Evolution of
the magnetic twist distribution as
shown in \Fig~\ref{fig4}d indicates that the reconnection adds
magnetic twist to the MFR~\citep{WangW2017}. The twist is added to the outer layer of
the rope, while its center keeps the original value of the
pre-flare configuration.  The rising path of the MFR deviates from the
vertical axis to the east side (the $-x$ direction), where
the magnetic pressure is weaker than the west side (the $+x$
direction).

In 3D, the MFR's surface (or boundary) is complex, but an arched tube
structure can be seen from the QSLs~(Supplementary Movie~1), within
which the magnetic twist is distinctly stronger than that of the
ambient flux~(Supplementary Movie 2, \Fig~\ref{fig7}a and b). With the rising of the MFR body, its
conjugated legs are rooted in the far-side polarities P1/N1
without connection to P0/N0.
Consequently, the initial elongated distribution of magnetic twist along the main PIL
becomes coherent in the two feet of the rope, which expand in size with time (Supplementary Movie 4).
Both feet are rather irregular: the south one forms closed rings
while the north one is even more complex and is split into two
fractions due to the mixed magnetic polarities there. As can be seen in Supplementary Movie~4,
the south foot expands from the initial closed QSLs that separate the P1-N1 flux
with other connections. Thus the initial P1-N1 flux provides a seed
for the subsequent erupting rope, and the weakly twisted flux remains
in the center of the rope. Evolution of the magnetic field lines traced
on the surface of rope, which forms the tube-like QSLs, demonstrates
the expansion of the rope in 3D (\Fig~\ref{fig5}b). This expansion is
also reflected in the observation of an EUV hot channel by SDO/AIA~94~{\AA}. As
observed from a nearly top view (see \Fig~\ref{fig5}c), there are
bright edges expanding from two sides of the flare ribbon (see also
the Supplementary Movie~3), which agrees well with the expanding surface of
the simulated rope (as seen in the same view angle, \Fig~\ref{fig5}b). Such EUV hot channel
is often associated with erupting MFR~\citep{ZhangJ2012, Cheng2012}, and here it is suggested to
 correspond specifically to the surface of MFR. There
appears no writhe of the rope, as seen in both the simulation and the observation,
indicating that the KI did not
occur. Thus the TI is probably the driving mechanism of the
eruption.

At the bottom surface, the QSL initially coinciding with the main PIL
bifurcates in two parallel ones which depart
from each other (see \Fig~\ref{fig7}a and
Supplementary Movie~4). This is due to the formation of the HFT below the MFR and the
continuous reconnection in the corona. This reconnection in 3D occurs in a
tether-cutting-like configuration (as illustrated in \Fig~\ref{fig6}b and c), which
cuts the connection of the twisted/sheared arcades at their inner
footpoints with the photosphere. In observation, the feet of erupting
MFR can be indicated by transient coronal dimming accompanied with eruptions,
which is thought to be resulted by the plasma evacuation along the legs of the erupting
MFR~\citep{Qiu2007, Webb2000}. In \Fig~\ref{fig6}d, a time difference of 304~{\AA} images
before and after the eruption is plotted, which distinctly show two
patches of coronal dimming in the polarities N1 and P1,
respectively. These dimming sites coincide well with the two feet of
the erupting MFR in the MHD model.

Our simulation reproduced the changing geometry of flare ribbons
(\Fig~\ref{fig7}), which correspond to the footpoints of magnetic
field lines that are undergoing reconnection. These field lines,
forming the surface of the MFR (or the tube-like QSLs), pass through
the CS (or the HFT) below the rising rope. As can be seen in
\Fig~\ref{fig7}d and Supplementary Movie~5, there are two main flare ribbons (the brightest
ones) along the main PIL, which exhibit a separation motion with
time. The footpoints of reconnecting field lines match strikingly well
the spatial location of the main double ribbons as well as their
separation.  In addition to the main double ribbons, there are
relatively weak ribbons extending in the two far-side polarities,
P1 and N1, as shown by the arrows in Fig~\ref{fig7}d. It can be seen that
these weak ribbons form a nearly closed ring in the south polarity, P1, while the
north one is rather complex. These ribbons should be produced by reconnection-released
energy reaching the far-end footpoints of the reconnecting field
lines traced from the CS, which forms the boundaries of the MFR's legs.
With the separation of the main double ribbons, the closed
region formed by the secondary ribbons also expands. This corresponds
to the process of more and more flux joining in the MFR through the
reconnection in the CS, which is basically in line with the 3D
standard flare model. In Supplementary \Fig~1, the separation speed of the two main ribbons from the simulation is compared with that from observation: the modeled one is $50 \sim 70$~km~s$^{-1}$ which is approximately $3\sim 5$ times larger than the observed one ($10\sim 20$~km~s$^{-1}$).
The reconnection rate, as measured by the ratio of inflow speed to the local Alfv{\'e}n speed (i.e., the inflow Alfv{\'e}n Mach number), is $0.05  \sim 0.1$ in the MHD simulation, which is comparable to the estimated values from direct observation analysis of different flares~\citep{SuY2013, SunJ2015}. Thus, the real value of the reconnection rate for this X9.3 flare should be $0.01\sim 0.03$, if we apply the same scale factor, i.e., the ratio of the simulated ribbon separation speed with the observed one, to our simulated reconnection rate.


Finally, the modeled post-flare loops, which is the short magnetic
arcades immediately below the flare CS, show a strong-to-weak
transition of magnetic shear with the progress of the flare
reconnection (Supplementary \Fig~2a). Such configuration evolution of flare
loops is well observed for typical two-ribbon
flares~\citep{Aulanier2012}. Another well-documented fact we have
reproduced (Supplementary \Fig~2b) is a permanent enhancement of the
transverse field along the main PIL on the bottom surface after
flare~\citep{WangH2012}.


\section*{Discussion}

We have revealed the topological evolution of magnetic configuration associated with
a solar flare through
a combination of observation data and numerical simulation.
Reconstruction of the coronal magnetic
field immediately prior to the flare 
results in an MFR, which is a basic building block of the standard
flare model. Much more complicated than any theoretical or idealized model,
the MFR consists of multiple bundles of field lines with
different connections and twist degrees. Owing to a strongly distorted, quadrupolar magnetic system as observed on the photosphere, the main body of the MFR
forms a C shape. This is unlike a typically observed coronal sigmoid that is often associated with pre-flare MFR.
Magnetic field lines of the MFR run horizontally over the strongly sheared PIL in
the core of the AR. The bottom of the rope is attached on the photosphere, resulting in BPs along the PIL.
Analysis of the decay index of the background
potential field in the vicinity of the MFR shows that a major part of
the MFR already enters into the TI domain.

The unstable nature of the pre-flare magnetic field results in
fast expansion and rising of the MFR in the MHD
simulation as initialized by the reconstruction data.
In the wake of the rising MFR, an HFT comes into being, i.e., an intersection of the QSLs that warp the rope. Strong current density thus accumulates in the HFT, forming a CS and reconnection is consequently triggered there. Magnetic twist is sequentially built up on the outer layer of the rope through reconnection of the field lines there.
The modeled magnetic configuration and evolution are found to be consistent with observed EUV features of the eruption, such as the expanding hot channels that are presumably the QSLs of the rope, the dark cavity with bright edge that corresponds to the cross section of the rope, and the coronal dimming in the feet of the rope. Most importantly,
by tracing the
newly-reconnected field lines from the CS to the bottom surface, we
have reproduced the location of two main flare ribbons as well as their separation with time, which is achieved for the first time.
Furthermore, the average reconnection rate of the flare could be estimated from the comparison of the simulated ribbon separation speed and the observed one. In addition to the main ribbons,
there are relatively weak or secondary flare ribbons extending to the
two feet of the rising rope, which 
actually correspond to the footpoints of the QSLs on the rope
surface. The
areas enclosed by these ribbons increase gradually as increasing amount of flux
joins the MFR through the reconnection.

In summary, without any artificial assumption on the magnetic configuration, we
reproduced the eruptive process of a great solar flare with numerical MHD simulation
based directly on the observed magnetogram. The recreated flare process is found to be in
correspondence with the standard flare model in 3D, thus providing the strong
evidence supporting such a model.

\section*{Methods}

\noindent\textbf{Instrument and data.}
The SDO/AIA can provide a full-disk image of the Sun simultaneously in 6 EUV filters, including 171~\AA, 193~\AA, 211~\AA, 335~\AA, 94~{\AA}, and 131~\AA. The spatial resolutions of all these filters are $0.6$~arcsec and the cadences are 12 seconds. The vector magnetogram used for our coronal field extrapolation is taken by SDO/HMI~\cite{Schou2012HMI}. In particular, we used the data product of the Space-weather HMI Active Region Patch (SHARP)~\cite{Bobra2014}, which has resolved the 180$^{\circ}$ ambiguity by using the minimum energy method, modified the coordinate system via the Lambert method and corrected the projection effect. Furthermore, this flare was associated with an erupting filament. For checking the location of the filament, the H$\alpha$ data, with spatial resolution of 1~arcsec and cadence of 1~minute, from Global Oscillation Network Group (GONG) are used as well.

All the data used in current study are publicly available: the SDO/HMI vector magnetograms and SDO/AIA images can be downloaded from the Joint Science Operations Center (JSOC) website \url{http://jsoc.stanford.edu/}; the GONG data can be downloaded from \url{https://gong2.nso.edu/}.

\bigskip

\noindent\textbf{Coronal Magnetic Field Reconstruction.}
The pre-flare coronal magnetic field is extrapolated by the
CESE--MHD--NLFFF code~\citep{Jiang2013NLFFF}. It belongs to the class
of magneto-frictional methods~\citep{Roumeliotis1996} that seek nonlinear force-free equilibrium
\begin{equation}\label{nlfff}
  (\nabla \times \vec B)\times \vec B = \vec 0
\end{equation}
for given
boundary value of observed vector magnetograms. It solves a set of
modified zero-$\beta$ MHD equations with a friction force using an
advanced conservation-element/solution-element (CESE) space-time
scheme on a non-uniform grid with parallel
computing~\citep{Jiang2010}. Starting from a potential field
extrapolated from the vertical component of the vector magnetogram,
the MHD system is driven to evolve by incrementally changing the
transverse field at the bottom boundary to match the vector
magnetogram, after which the system will be relaxed to a new
equilibrium. The code has an option of using adaptive mesh refinement
and multi-grid algorithm for optimizing the relaxation process. The
computational accuracy is further improved by a magnetic-field
splitting method, in which the total magnetic field is divided into a
potential-field part and a non-potential-field part and only the
latter is actually solved in the MHD system. Before inputting to the
code, the vector magnetograms are usually required to be preprocessed
to reduce the Lorentz force involved. Furthermore, to be
consistent with the code, we developed a unique preprocessing
method~\citep{Jiang2014Prep} that also splits the vector magnetogram
into a potential part and a non-potential part and handles them
separately. Then the non-potential part is modified and smoothed by an
optimization method~\citep{Wiegelmann2006} to fulfill the
conditions of total magnetic force-freeness and
torque-freeness. Details of the code are described in a series of
papers~\citep{Jiang2013MHD, Jiang2014formation, Jiang2014NLFFF}. It is
well tested by different benchmarks including a series of
analytic force-free solutions~\citep{Low1990} and numerical
MFR models~\citep{Titov1999, Ballegooijen2004}, and have been applied to the SDO/HMI vector
magnetograms~\citep{Jiang2013NLFFF, Jiang2014NLFFF}, which enable to reproduce
magnetic configurations in very good agreement with corresponding
observable features, including coronal loops, filaments, and sigmoids.
\bigskip{}

\noindent\textbf{MHD Model.}
The MHD simulation is realized by solving the full set of 3D,
time-dependent ideal MHD equations with solar
gravity~\citep{Jiang2016NC}.  The initial condition consists of
magnetic field provided by the NLFFF model and a hydrostatic
plasma. The initial temperature is uniform, with a value typically in
the corona, $T=10^{6}$~K (sound speed $c_{S}=128$~km~s$^{-1}$). The
initial plasma density is uniform in horizontal direction and
vertically stratified by the gravity. To mimic the coronal low-$\beta$
and highly tenuous conditions, the plasma density is configured to
make the plasma $\beta$ less than $0.1$ in most of the
computational volume. The smallest value of $\beta$ is $ 5\times
10^{-4}$, corresponding to the largest Alfv{\'e}n speed $v_{\rm A}$ of
approximately 8~Mm~s$^{-1}$. The units of length and time in the model
are $L=11.5$~Mm (approximately 16~arcsec on the Sun) and
$\tau=L/c_{S}=90$~s, respectively. The MHD solver is the same CESE
code described in ref\citep{Jiang2010}. We use a non-uniform grid
adaptively based on the spatial distribution of the magnetic field and
current density in the NLFFF model. This grid is designed for the sake
of saving computational resources without losing numerical accuracy,
and more details of this can be found in ref~\citep{JiangC2017}. The
smallest grid is $\Delta x = \Delta y = 2\Delta z=0.36$~Mm
(approximately 0.5~arcsec on the Sun). A moderate viscosity $\nu$, which
corresponds to Reynolds number $R_e = L v_{\rm A} /\nu$ of $\sim 10^2$, is used to keep the
numerical stability of the code running for the whole duration of the flare eruption process. No explicit resistivity is
included in the magnetic induction equation, and magnetic reconnection
is still allowed due to numerical resistivity $\eta$, which corresponds to Lundquist number (or magnetic Reynolds
number) of $S = L v_{\rm A}/\eta $ of $\sim 5\times 10^3$.
Although there is no doubt that the viscosity and
numerical resistivity in our model overestimate the real values in the
coronal plasma (which are on the order of $10^8\sim 10^{10}$), the basic magnetic topological evolution as simulated
is still robust~\citep{Jiang2016NC}.  The simulation is stopped before
any disturbance reaches the numerical boundaries. At the bottom
boundary (i.e., the coronal base), all the variables are fixed except
the transverse components of the magnetic field, which are specified by
linear extrapolation from the two adjacent inner points along the $z$-axis.

In the combination of NLFFF model and MHD simulation, one should
note almost all the available NLFFF codes actually generate
non-force-free magnetic field data with residual Lorentz force that
is often non-negligible. This is indicated by the misalignment of the
electric current $\vec J$ and magnetic field vector $\vec B$~\citep{Schrijver2006},
which is usually measured by CWsin, a current-weighted sine of the
angle between $\vec J$ and $\vec B$. CWsin is typically in the range
of $0.2$ to $0.4$~\citep{Schrijver2008a, DeRosa2009,
  DeRosa2015}. Another metric $E_{\curl \vec B}$ measuring more
directly the residual Lorentz force is defined as the average ratio of
the force to the sum of the magnitudes of magnetic tension and pressure
forces~\citep{DuanY2017}. In the studied event here, these two metrics
for the CESE--MHD--NLFFF extrapolation are respectively, CWsin $=0.23$
and $E_{\curl \vec B} = 0.17$ in the strong current region. These
metrics are reasonably small as compared with other codes~(e.g.,
see the last column of Table~2 in ref~\cite{DeRosa2015}), but such residual force can instantly induce plasma motions in a low-$\beta$ and highly tenuous plasma environment. This motion provides a way of perturbing
the system. If the system is significantly far away from unstable,
such unbalance state will relax to a new MHD equilibrium as the
induced motion can alter the magnetic field, which in turn generates
restoring Lorentz force to stabilize the system. Else if the system is
unstable or not far away from unstable regime, the perturbation could grow
and lead to a drastic evolution of the system as driven by the instability. Thus the
combination of NLFFF and MHD model can be used to test the potentially unstable nature or
 instability of NLFFF.
Although such combination of NLFFF and MHD models might not be able to identify the true mechanism triggering flare, it still provides a viable tool to reproduce the fast magnetic evolution during the flare.

\bigskip
\noindent\textbf{Magnetic field analysis tools.}
A set of magnetic field analysis methods is used in this study, including search of BPs, calculation of magnetic twist number, squashing degree and decay index, which are described in the following.

BPs are places on photospheric PIL where the transverse field directs from the negative polarity to positive one. This is inverse to a normal case that transverse field directs from positive flux to negative one, and thus the field line is concave upward. BPs can be located by searching the point on the magnetogram where the conditions
\begin{equation}\label{eq1}
  \vec B\cdot \nabla B_z > 0,\ \ B_z=0
\end{equation}
are satisfied. Magnetic dips are searched using the same conditions but applied for the full 3D volume of the field.

The magnetic twist number $T_w$ for a given (closed) field line is defined by~\citep{LiuR2016}
\begin{equation}\label{Tw}
  T_w=\int_L \frac{(\nabla \times \vec B)\cdot \vec B}{4\pi B^2} dl
\end{equation}
where the integral is taken along the length $L$ of the magnetic field line from one footpoint to the other.

The squashing degree $Q$ is derived based on the mapping of two footpoints for a field line. Specifically, a field line starts at one footpoint $(x,y)$ and ends at the other footpoint $( X(x,y), Y(x,y) )$. Then the squashing degree associate with this field line is given by~\citep{Titov2002}
\begin{equation}
  \label{eq:Q}
  Q = \frac{a^{2}+b^{2}+c^{2}+d^{2}}{|ad-bc|}
\end{equation}
where
\begin{equation}
  a = \frac{\partial X}{\partial x},\ \
  b = \frac{\partial X}{\partial y},\ \
  c = \frac{\partial Y}{\partial x},\ \
  d = \frac{\partial Y}{\partial y}.
\end{equation}
Usually QSLs can be defined as locations where $Q>>2$.

In the torus instability (TI)~\cite{Kliem2006PRL}, which is a result of the loss of balance between the ``hoop force'' of the rope itself and the
``strapping force'' of the ambient field, the decay index $n$ plays a key role. It quantifies the decreasing
speed of the strapping force along the distance from the torus
center.
Here $n$ is calculated in the vertical cross section perpendicularly crossing the main axis of the rope~(see \Fig~\ref{fig3}c) in such manner: we regard the bottom PIL point, named O (denoted by the black circle in the figure) as the center of the torus, and for a given grid point P, $n({\rm P})=-d \log (B_{\rm p})/ d \log(h)$,  where $B_{\rm p}$ is the magnetic field component perpendicular to the direction vector $\vec r_{\rm OP}$, and $h = |\vec r_{\rm OP}|$. Here the strapping field is approximated by the potential field model that matches the $B_z$ component of the photospheric magnetogram~\citep{Aulanier2010, Jiang2013MHD}.

\bigskip

\newpage

\newpage
\section*{}
\subsection*{Acknowledgments}
Data from observations are courtesy of NASA {SDO}/AIA and the HMI
science teams. The computation work was carried out on TianHe-1 (A) at
the National Supercomputer Center in Tianjin, China.
This work is supported by the National Natural Science Foundation of China
(41574170, 41574171, 41531073, 41731067). C.J. and Q.H. acknowledge NSF grant AGS-1650854 for partial support.

\subsection*{Author contribution}
C.J. developed the model, performed the result analysis
and wrote the first draft. P.Z. performed part of the observation analysis. All authors participated in discussions and revisions on the manuscript.

\subsection*{Competing financial interests:}
The authors declare no competing financial interests.


\subsection*{Supplementary information}
Supplementary Fig 1 and Movies 1--5 are accompanied with this paper.

\noindent {\bf Supplementary Movie 1:} Rotating view of 3D structure of the QSLs that warp the erupting MFR at simulation time $t=1$. (a) The iso-surface of $Q=1000$, (b) Sampled magnetic field lines that form the QSLs. The colors denote the value of height $z$.

\noindent {\bf Supplementary Movie 2:} Slice through the volume showing the squashing degree $Q$ (a) and the magnetic twist degree $T_w$ (b).

\noindent {\bf Supplementary Movie 3:} Observation of the flare eruption in SDO/AIA~94~{\AA}.

\noindent {\bf Supplementary Movie 4:} Temporal evolution of the magnetic squashing degree (a) and twist degree (b) on the bottom surface.

\noindent {\bf Supplementary Movie 5:} Observation of the evolution of flare ribbons in SDO/AIA~304~{\AA}.


\newpage

\begin{figure}[h]
  \centering
  \includegraphics[width=\textwidth]{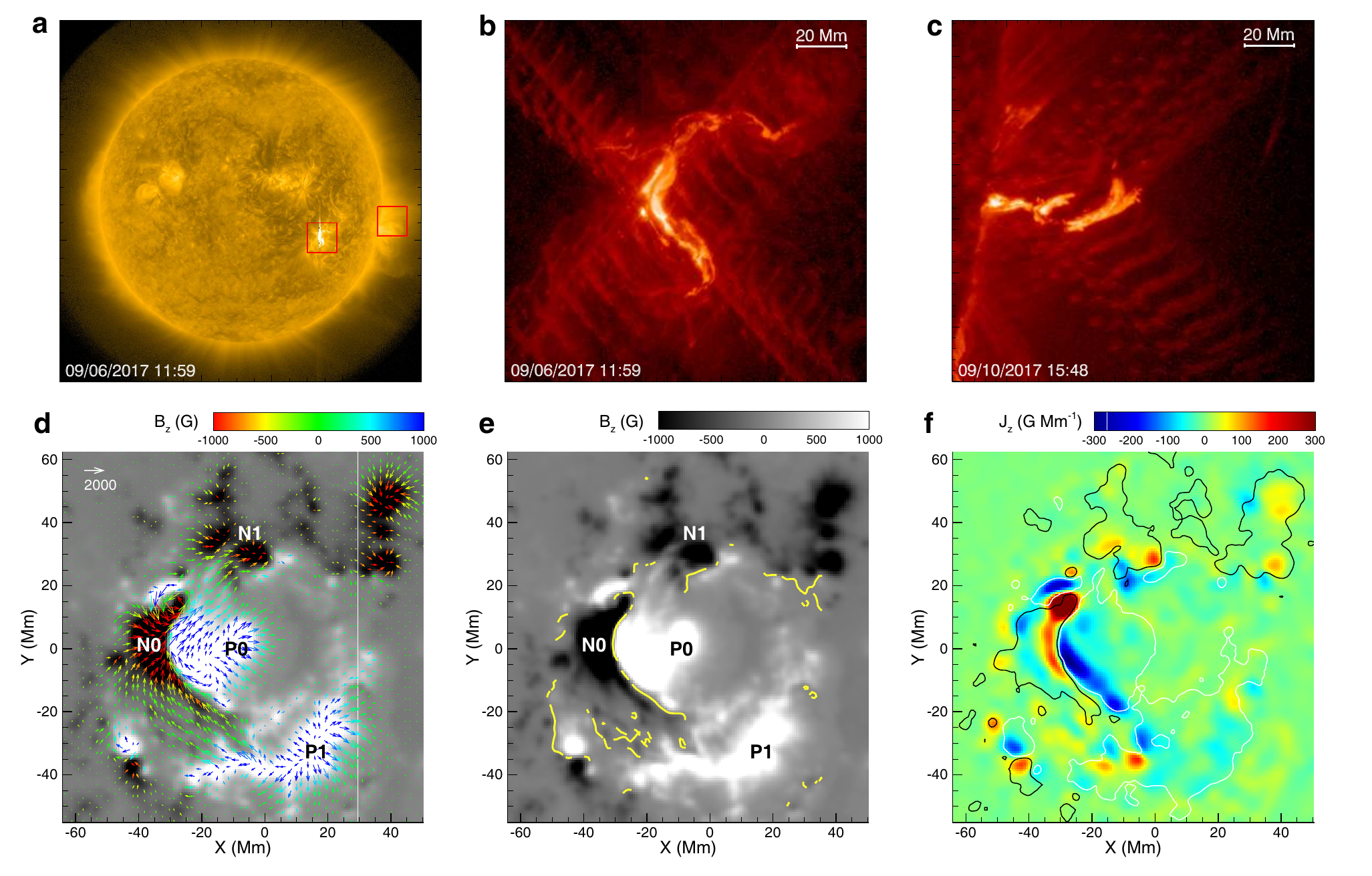}
  \caption{
  \textbf{The flare location and photospheric magnetic field.}
  (a) Full-disk image of the Sun observed in SDO/AIA 171~{\AA}. The two boxes indicate the locations of the on-disk X9.3 flare on September 6 and the limb X8.2 flare occurred on September 10. (b)-(c) SDO/AIA 304~{\AA} image of the X9.3 flare and the X8.2 flare, respectively.
  (d)-(f) SDO/HMI vector magnetogram taken at 11:36~UT on September 6, which
  is 17~min before the X9.3 flare onset. In (d), the magnetic flux distribution, i.e., $B_z$, is overlaid by the transverse field vector $(B_x, B_y)$ as denoted by the colored arrows. The main magnetic polarities P0, N0, P1, and N1 are
  labeled. In (e) the yellow curves are the BP locations along the PIL. (f)
  Distribution of the vertical current density, which is defined as $J_z = \partial_x B_y - \partial_y B_x$. The contour lines are plotted for $B_z = -500$~G (colored as black) and 500~G (white). The ratio of the direct current (DC) to the return current (RC) for the positive flux is $|{\rm DC}/{\rm RC}|^+ = 2.31$, and for the negative flux is $|{\rm DC}/{\rm RC}|^- =2.26$.}
\label{fig1}
\end{figure}

\begin{figure}[h]
  \centering
  \includegraphics[width=\textwidth]{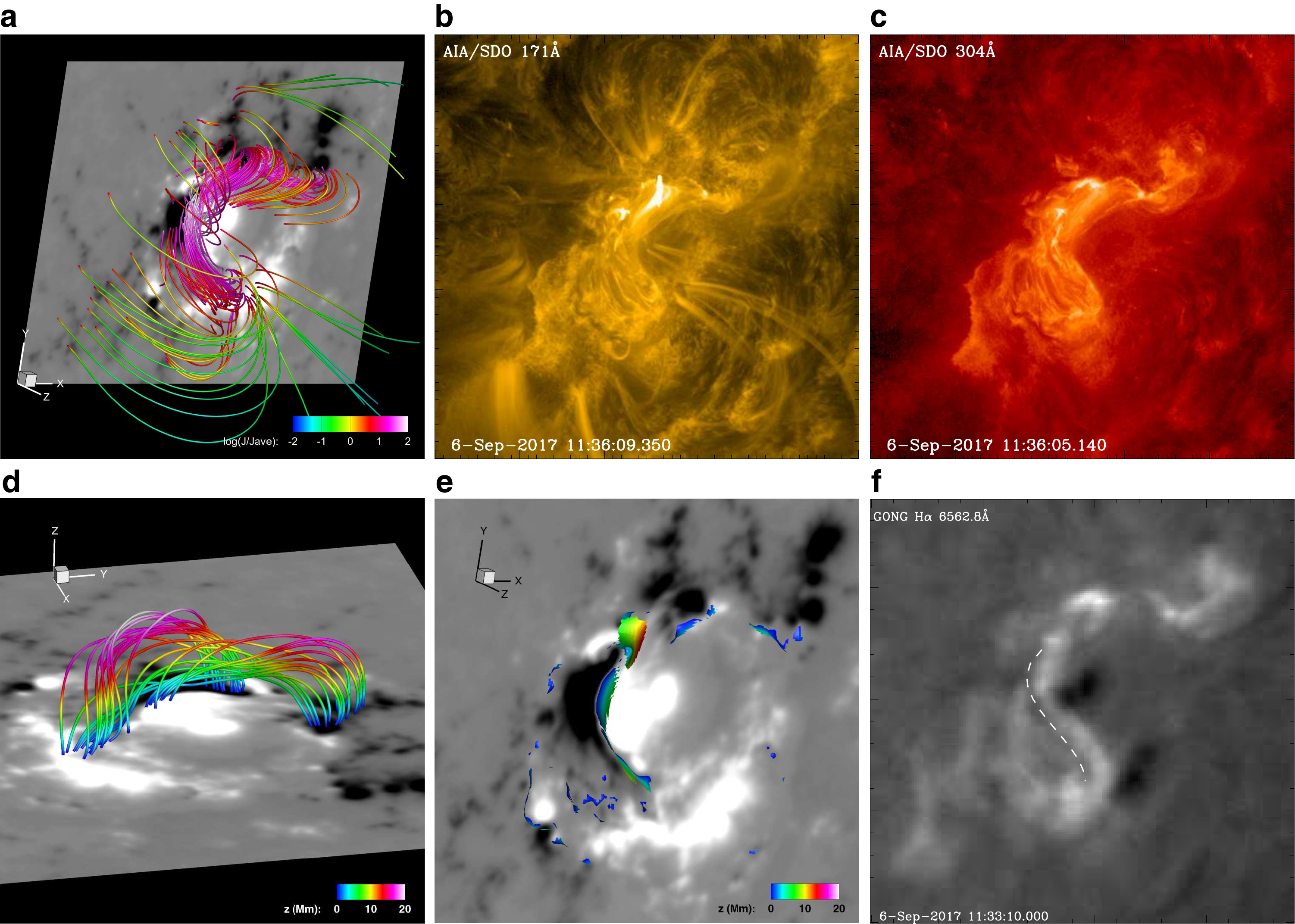}
  \caption{
  \textbf{Comparison of the reconstructed magnetic field with the observed features of the solar corona prior to the flare.}
    (a) SDO view of sampled magnetic field lines of the NLFFF reconstruction. The color of the lines represents the value of current density $J$ (normalized by its average value $J_{\rm ave}$ in the computational volume). The background is the photospheric magnetogram.
    (b) and (c) SDO/AIA 171~{\AA} and 304~{\AA} images of the pre-flare corona.
    (d) The low-lying magnetic field lines in the core region. The field lines are color-coded by the value of height $z$. (e) Locations of dips in the magnetic field lines, and the color indicates the value of height $z$.
    (f) GONG H$\alpha$ image of the AR. The dashed curve denotes the location of a long filament.}
  \label{fig2}
\end{figure}

\begin{figure}[htbp]
  \centering
  \includegraphics[width=\textwidth]{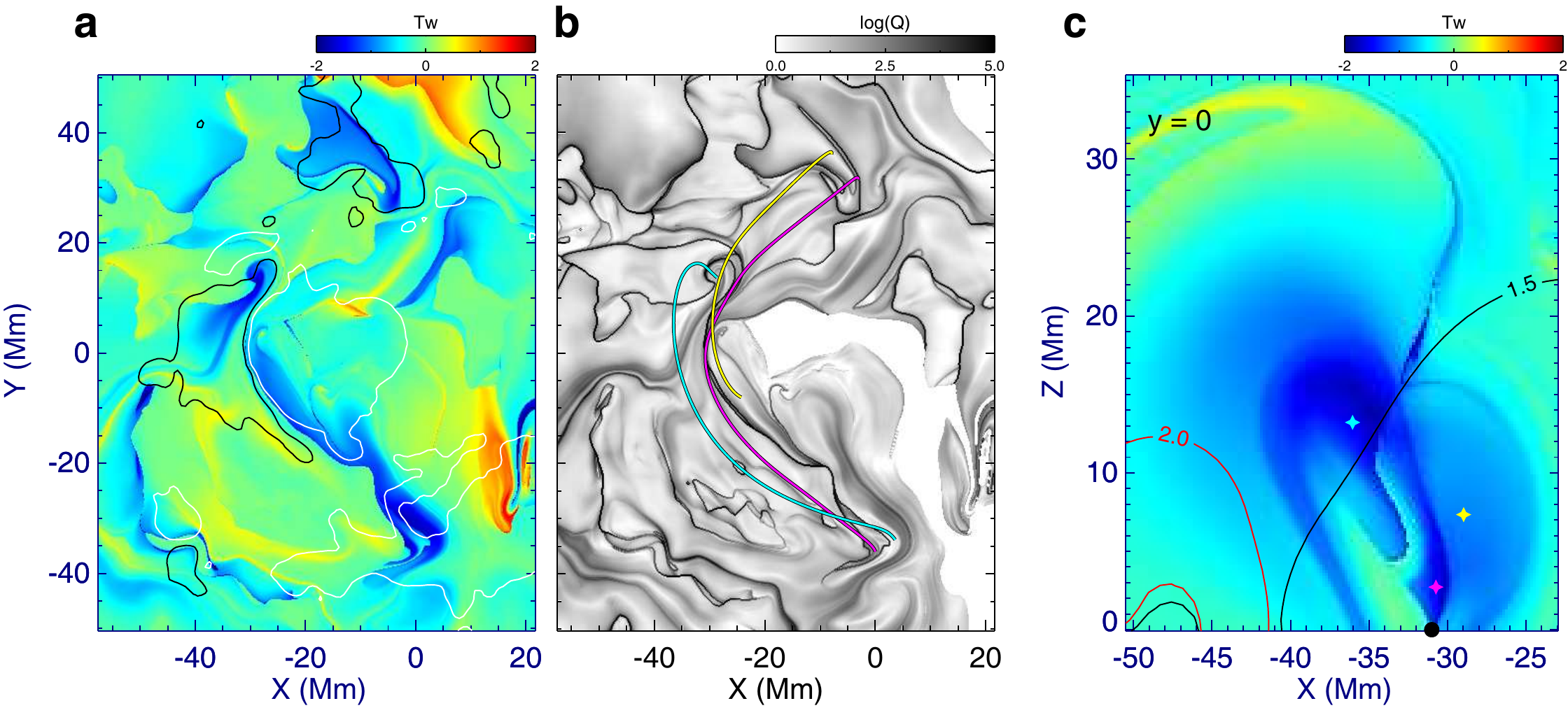}
  \caption{
  \textbf{Detailed configuration of the reconstructed pre-flare magnetic field.}
  (a) Map of magnetic twist number $T_w$ at the bottom surface $z=0$. Overlaid are contour lines for $B_z=500$~G (white) and 500~G (black).
  (b) Map of magnetic squashing factor $Q$ at the bottom. The black thin lines as formed by the large-$Q$ value are locations of magnetic topology separatries and QSLs where the magnetic field-line mapping is
  discontinuous or steeply changes. Three field lines with different colors are plotted to represent the magnetic flux of different connections, which make up the MFR.
  (c) Twist number distribution on a vertical cross section ($y=0$). The three colored stars denote the intersection points of the sample field lines shown in (b) with the cross section. The black circle indicates the main PIL. The contour lines are shown for the decay index $n=1.5$ and 2 (see Methods).}
  \label{fig3}
\end{figure}

\begin{figure}[htbp]
  \centering
  \includegraphics[width=\textwidth]{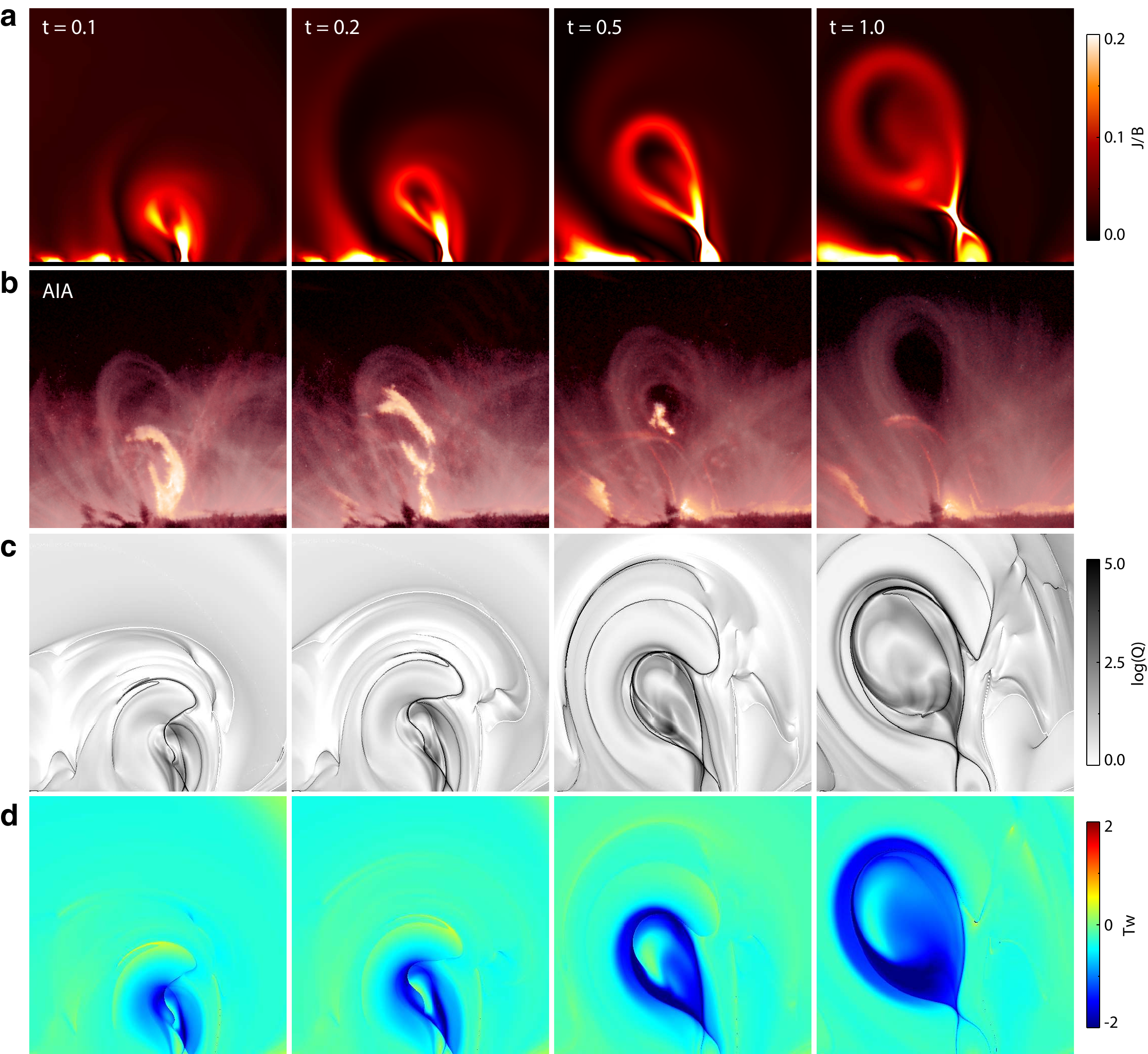}
  \caption{
  \textbf{Temporal evolution of the eruptive structure in 2D view.}
  (a) Distribution of current density on the vertical cross section (the $y=0$ plane). Here the current density is normalized by local magnetic field strength, which provides a high contrast of thin current layers with other volumetric currents.
  (b) SDO/AIA images of the X8.2 flare observed at the solar limb. The images are made by combination of two AIA channels 211~{\AA} and 304~{\AA}, and they are rotated to roughly match the direction of the simulated eruption.
  (c)-(d) Distributions of magnetic squashing degree $Q$ and twist number $T_w$, respectively, on the same cross section in (a).}
  \label{fig4}
\end{figure}

\begin{figure}[htbp]
  \centering
  \includegraphics[width=\textwidth]{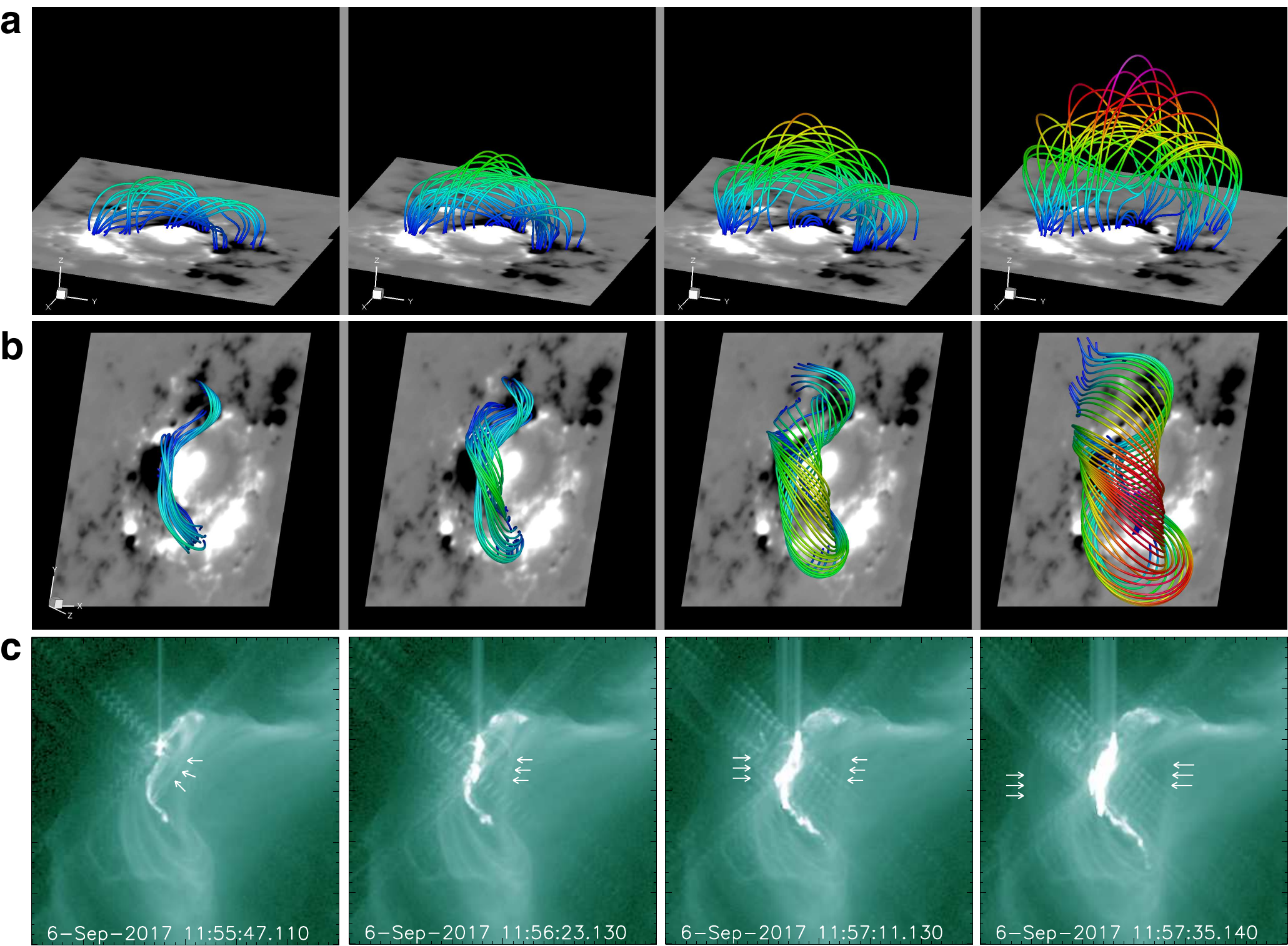}
  \caption{
  \textbf{The eruptive structure in 3D and comparison with SDO/AIA observation.}
  (a) Side view of sampled magnetic field lines of the erupting MFR. The magnetic field lines are false-colored by the value of height $z$ for a better visualization. The bottom surface is shown with the photospheric magnetogram.
  (b) Magnetic field lines that form the surface of the MFR (see Supplementary Movie 1 for a 3D rotation view of these field lines). The view angle is arranged to be the same as that of the SDO.
  (c) SDO/AIA~94~{\AA} observations of the erupting process. Two sets of arrows mark the two expanding edges, presumably corresponding to the expanding surface of the MFR. Such expanding features can be seen more clearly in the Supplementary Movie~3.}
  \label{fig5}
\end{figure}

\begin{figure}[h]
  \centering
  \includegraphics[width=\textwidth]{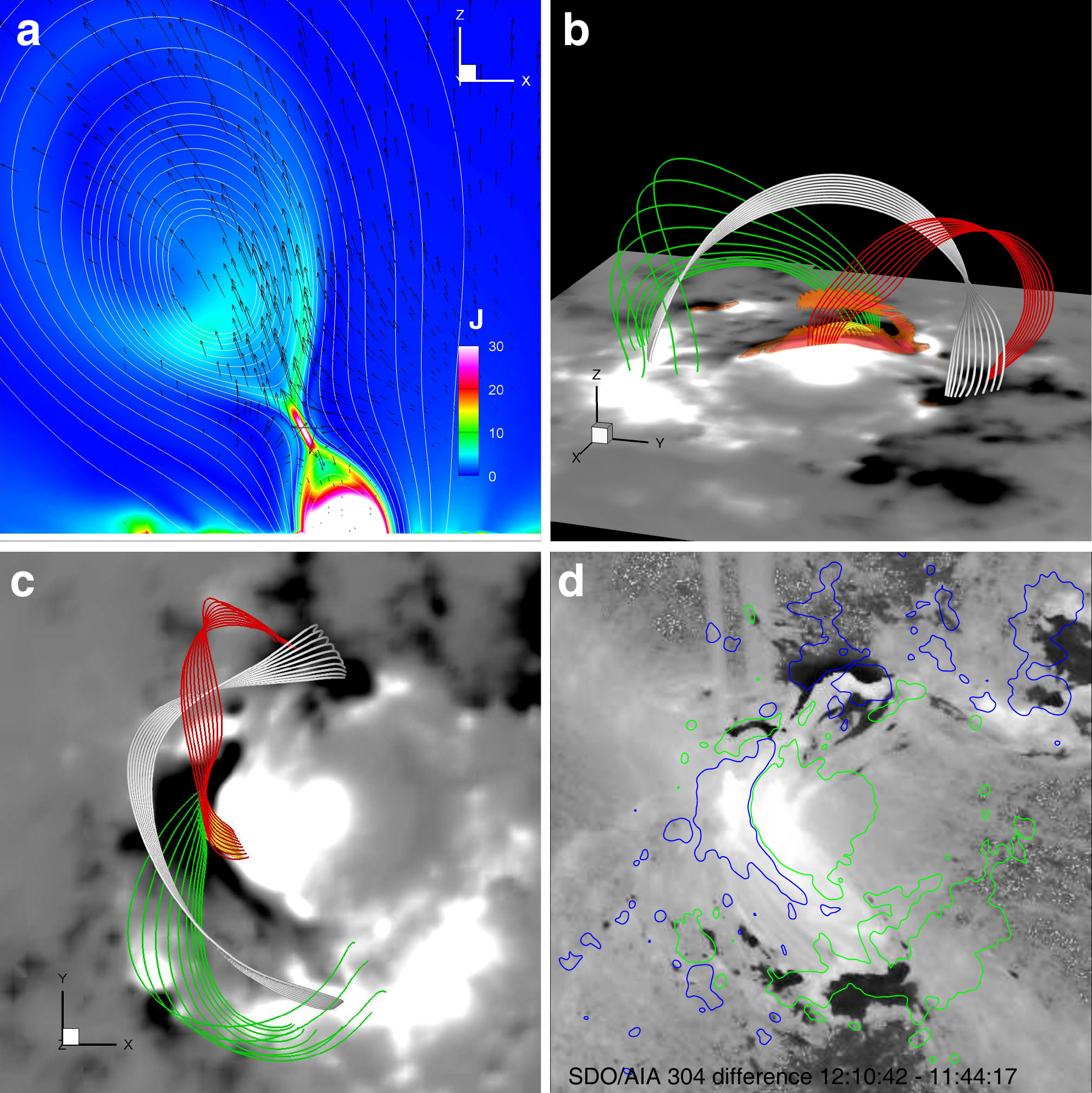}
  \caption{
  \textbf{Illustration of the reconnection process below the rising
    MFR.} (a) Current density distribution
    and plasma flows (denoted by arrows) on the vertical cross section (the $y=0$ plane) at time of $t=1.0$. The white lines are 2D field lines tracing
    on the plane. (b) 3D configuration of the reconnection.
    The white lines represent the main body of the
    MFR. The red and green lines represent the reconnecting
    field lines below the rope.
    Their inner footpoints are sheared past each along the PIL,
    and thus the field directions change abruptly across the CS, in which reconnection takes
    place, results in a long field line joining the MFR and a short
    arcade below which forms the post-flare loops (as shown by the yellow lines). The objects colored in
    red and orange are thin layers with strongest current density throughout the volume, showing the CS in 3D.
    (c) Top view of the same magnetic field lines shown in (b).
    (d) Difference of AIA 304~{\AA} images of post-flare with pre-flare time showing two dimming sites (or transient coronal holes) that match the locations of the MFR legs. The contour lines are shown for $B_z = -500$~G (colored in blue) and $500$~G (colored in green).}
  \label{fig6}
\end{figure}

\begin{figure}[htbp]
  \centering
  \includegraphics[width=\textwidth]{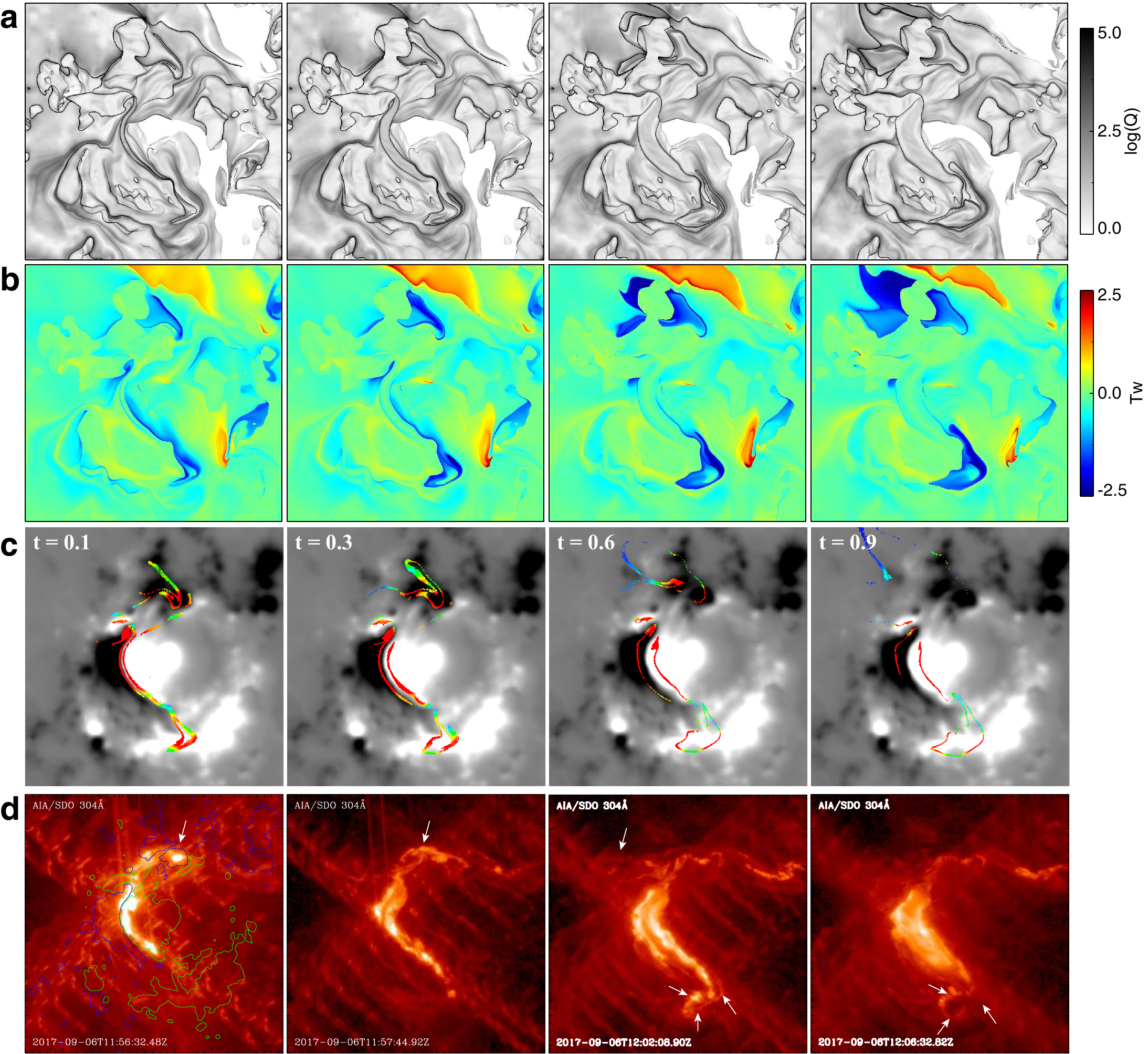}
  \caption{
  \textbf{Structures and evolution at the bottom surface.}
  (a) Magnetic squashing degrees. (b) Magnetic twist numbers.
  (c) Magnetic footpoints (color dots) of the field lines that are traced from the CS to the bottom surface. The colors represent the strength of local magnetic field, red for strong and blue for weak. (d) SDO/AIA~304~{\AA} images of the flare ribbons. The arrows denote the two weak ribbons that form in the far-side polarities P1 and N1.}
  \label{fig7}
\end{figure}

\begin{figure}[htbp]
  \centering
  \includegraphics[width=\textwidth]{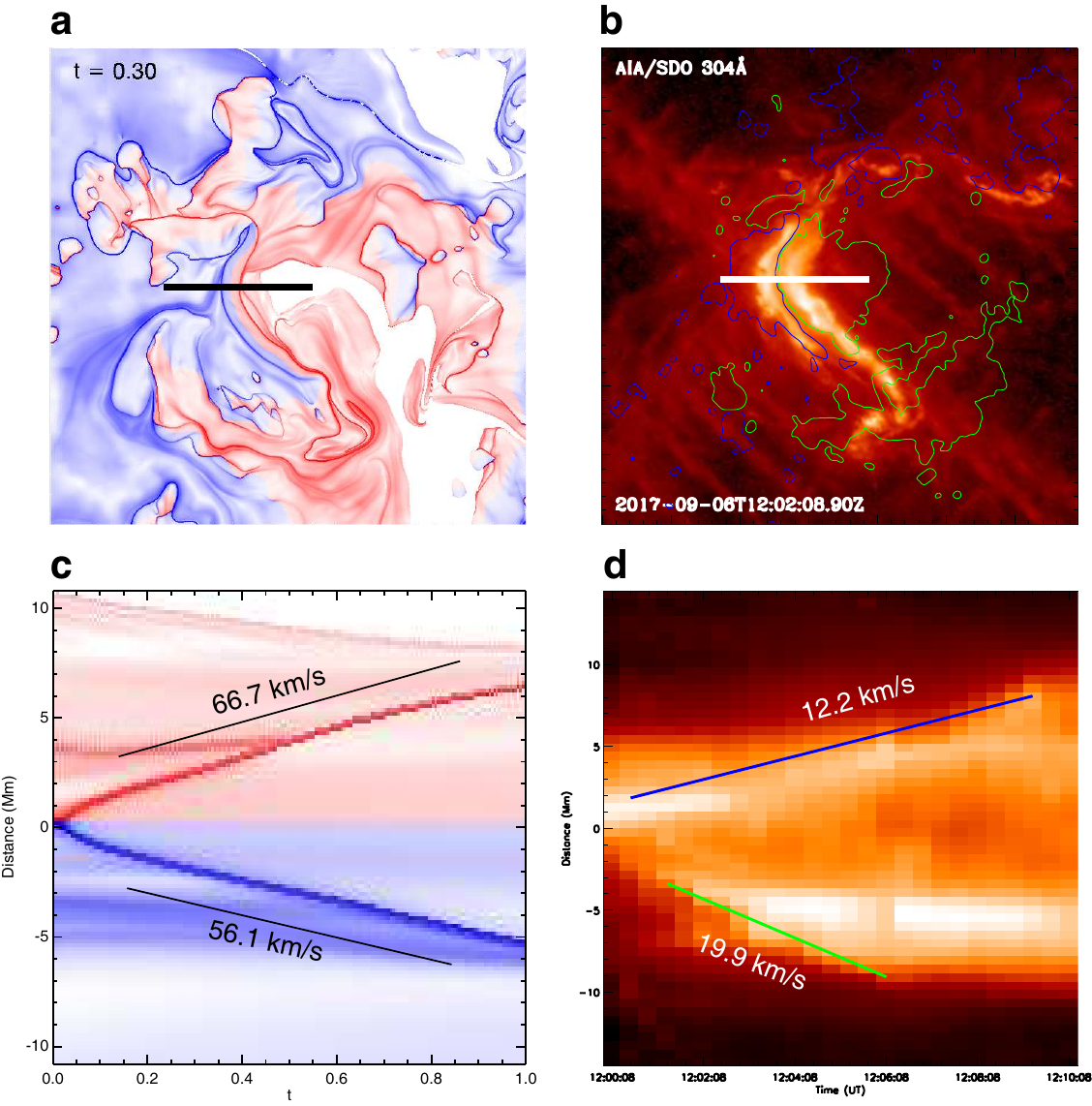}
  \caption*{
  \textbf{Supplementary Fig 1: Comparison of flare-ribbon separation speed from MHD simulation and that from observation.}
  (a) The QSL map on the bottom at time of $t=0.3$. The black line segment denotes the location where a time sequence of stack is plotted in (c) for tracking the separation of the main QSLs that maps the reconnection footpoint. (b) SDO/AIA 304~{\AA} image of the flare ribbons. In the same way, the white line segment is the location of the time stack shown in (d), which shows the separation motion of the main flare ribbons. The lines shown in (c) and (d) represent approximately the speeds of the ribbons with time, respectively.}
  \label{Sfig1}
\end{figure}

\begin{figure}[htbp]
  \centering
  \includegraphics[width=\textwidth]{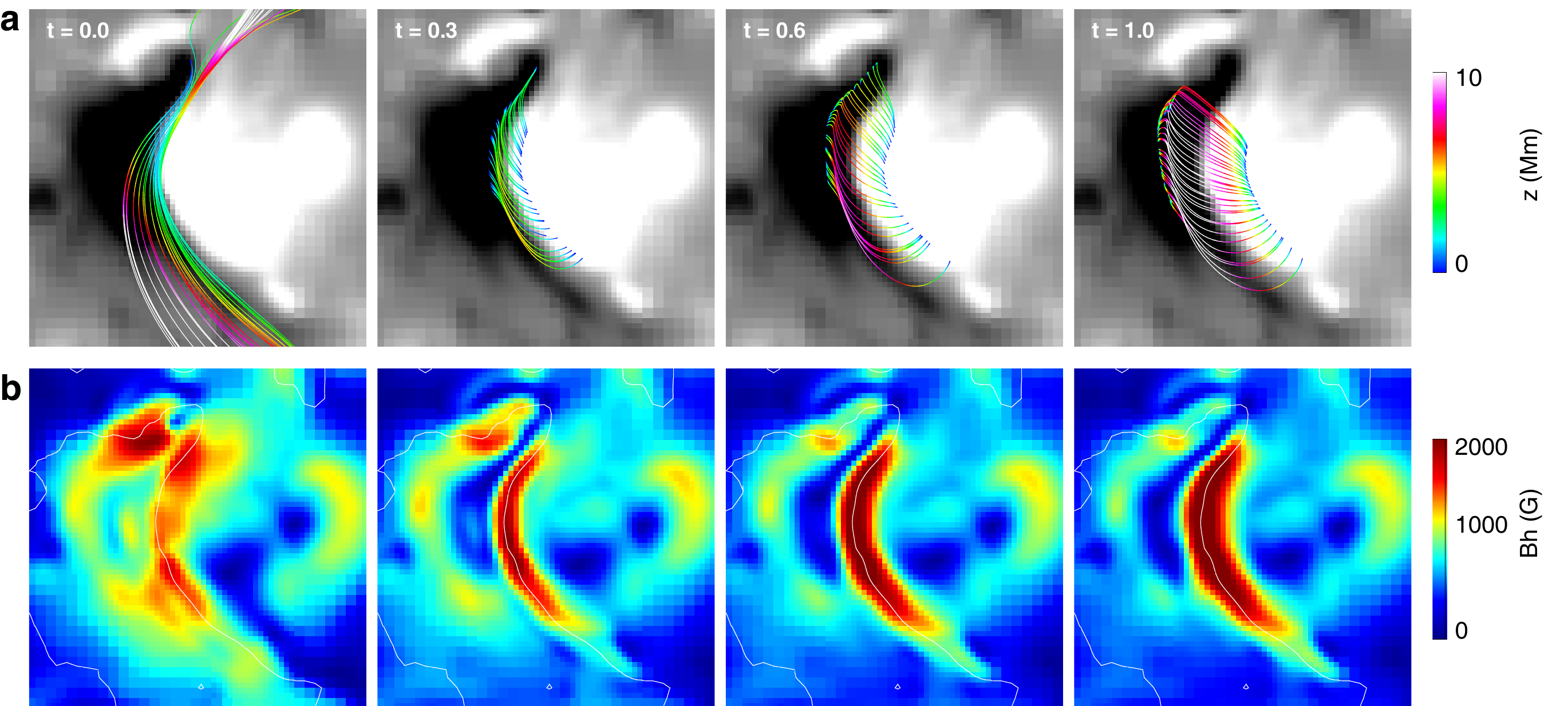}
  \caption*{
  \textbf{Supplementary Fig 2: Strong-to-weak shear transition in post-flare loops and enhancement of horizontal field.}
  (a) The closed magnetic arcades immediately below the reconnection site. Note that initially ($t=0$) they are long field lines that form the pre-flare MFR. The colors of the lines denote the value of height $z$. The background is shown with $B_z$ on the bottom. (b) Distribution of the horizontal field $B_h = \sqrt{B_x^2+B_y^2}$ on the bottom. The PIL of the magnetic flux $B_z$ is shown by the white-colored lines.}
  \label{fig8}
\end{figure}

\end{document}